\documentclass[11pt]{article}
\usepackage[english]{babel}
\input{epsf}
\usepackage{mathtext}
\usepackage{graphics,dcolumn,bm,float}
\usepackage{amssymb,amsmath,rotate,color}
\usepackage{times}
\newcommand{\hhh}{{\cal H}}

\newcommand{\CN}{{\cal N}}
\newcommand{\be}{\begin{equation}}
\newcommand{\ene}{\end{equation}}
\newcommand{\ba}{\begin{array}}
\newcommand{\ea}{\end{array}}

\newcommand{\bsigma}{\mbox{\boldmath$\sigma$}}

\begin{document}

\title{Andreev reflection and subgap conductance in monolayer $MoS_2$ ferromagnet/$s$ and $d$-wave superconductor junction}
\author{H. Goudarzi\footnote{Corresponding author: h.goudarzi@urmia.ac.ir ; goudarzia@phys.msu.ru}, M. Khezerlou\footnote{m.khezerlou@urmia.ac.ir}, S. F. Ebadzadeh \\
\footnotesize\textit{Department of Physics, Faculty of Science, Urmia University, Urmia, P.O.Box: 165, Iran}}
% Remove command to get current date 
\date{}
\maketitle

\begin{abstract}
The accurate and proper form of electron-hole excitations and corresponding Dirac-like spinors of monolayer molybdenum disulfide superconductor are exactly obtained. Andreev reflection and resulting subgap conductance in a $MoS_2$-based ferromagnetic superconducting (F/S) junction is accurately investigated in terms of dynamical characteristics of system. Due to spin-splitting energy gap in the valence band and nondegenerate $K$ and $K'$ valleys, the ferromagnetic exchange energy $\sigma h$ can cause a distinct behavior of Andreev process between spin-up and spin-down charge carriers belonging to different valleys. The chemical potential is necessarily fixed by a determined range in order to occur the retro Andreev reflection. Given one-particle superconducting bispinors enable us to explicitly involve the anisotropic superconducting gap $\Delta_S$ under electron-hole conversion, i.e., taking place in $d$-wave pair coupling. The effect of such gap is exactly explained in terms of the dependence of the Andreev process on the electron incidence angle at the interface.
\end{abstract}

\section{Introduction}

Two-dimensional condensed matters such as graphene $\cite{NG}$ and monolayer molybdenum disulfide (ML-MDS) $\cite{ML,KS,RR}$ including Dirac-like charge carriers can present itself as capable structures to observe distinct transport properties resulted from Andreev reflection (AR) and Klein transmission. By the Blonder-Tinkham-Klapwijk $\cite{BTK}$ formalism, the peculiar Andreev process results in a finite conductance in a normal/superconductor junction at the electron excitations below the superconducting gap $\Delta_S$. Another interesting feature of AR has been proposed by Beenakker $\cite{B}$ as specular AR, when a N/S proximity junction is realized in graphene, where an electron from the conduction band is reflected as a hole in the valence band, in which the reflection angle is inverted with respect to the incidence. This can be controlled by the bias voltage (electron excitation) dependence of the subgap Andreev conductance. Recently, the AR was studied at the interface of ML-MDS superconductor/normal metal $\cite{MRA}$, where the authors show the p/n-doping effect (the magnitude of the chemical potential in the normal region relative to superconductor region) on the retro AR. This may gain more attention, since charge carriers exhibit either electron-like or hole-like quasiparticles belonging to two inequivalent nondegenerate $K$ and $K'$ valleys. Comparing with graphene, such attention for ML-MDS is highlighted by some distinct features of $MoS_2$: i) existence of direct band-gap in low-energy band structure in the visible frequency rang ($\approx 0.95\:eV$), ii) strong spin-orbit coupling (SOC) resulted from heavy metal atoms, iii) breaking the valley degeneracy with a valley-contrasting spin splitting ($\approx 0.1-0.5 \:eV$) caused by inversion symmetry breaking $\cite{RR,ZC,XL}$. In this paper, firstly we, in particular, investigate the explicit dependence of Andreev process on the electron incidence angle in the F/S and N/S junctions by determining the allowed chemical potential of ferromagnetic or normal region due to the significant spin-splitting of the valence band in $MoS_2$. To do this, we obtain the explicit expression of the ML-MDS superconducting electron-hole excitations and corresponding Fermi wavevectors, which enables us to find the exact and appropriate form of Dirac-Bogoliubov-de Gennes (DBdG) spinors. We show that these spinors are fundamentally different from those obtained in the previous works $\cite{MRA,RMZ}$, so that we are allowed to consider the difference of superconducting gap under electron-hole converting, taking place in the $d$-wave superconductivity $\cite{L,T,IL}$. Secondly, we focus on the F/S structure, because of the exchange splitting energy $h$ of F metal may induce a large spin-splitting of $K'$ valley in the valence band, which results in a novel behavior of pseudo-relativistic Klein tunneling giving rise to a tunneling conductance difference between spin-up and spin-down carriers and resulting magnetoresistance $\cite{KG}$. The same structure with graphene was studied in $\cite{M}$.

However, proximity-induced superconductivity and ferromagnetism in the $MoS_2$ can be experimentally achieved $\cite{G,TMS,Y,R,ZS,LZ,MW,MG,TV,MZP}$. Recently, the physics of spin and valley coupling $\cite{XL}$ and ferromagnetic/superconductor/ferromagnetic $\cite{MA}$ junction have been studied in the ML-MDS structures.
Further, the contribution of Schrodinger-like terms (topological and difference mass between electron and hole terms) $\cite{RMA}$ are taken into account to the Dirac-like one-particle superconductor excitations. Moreover, we investigate the effect of anisotropic $d$-wave pairing energy gap $d_{x^2-y^2}$ in the Hamiltonian of ML-MDS and the resulting AR, because the sign of pair potential $\Delta_{S}^{e(h)}=\Delta_0\cos(2\theta_s^{e(h)}-(+)2\alpha_s)e^{i\phi}$ may be changed by electron-hole conversion. This leads to form the zero-energy states in the similar Josephson junction $\cite{TB}$ and corresponding zero-bias Andreev conductance. Obtained by us superconducting wavefunctions allow to explicitly exert this feature in the quasiparticle states. This paper is organized as follows. Sec. 2 is devoted to present the proposed model and formalism to obtain the exact form of $MoS_2$ superconducting dispersion energy and corresponding spinors. The normal and Andreev reflection coefficients are found by matching the wavefunctions at the interface. The numerical results of AR and resulting tunneling conductance considering the strong spin-valley effect caused by ferromagnetic exchange field and also asymmetric superconducting order are presented, and their main characteristics are discussed in sections 3 and 4. Finally, we close with a brief summary in Sec. 5.

\section{Theoretical formalism}

A typical $F/S$ structure on top of a ML-MDS sheet is introduced with the configuration that the ferromagnetic and superconductor regions are extended from $x=-\infty$ to $x=0$ and from $x=0$ to $x=+\infty$ for all $y$, respectively. The low-energy band structure of ML-MDS can be described by the modified Dirac Hamiltonian. This Hamiltonian in addition to the first order term of momentum for 2D massive fermions, contains the quadratic terms originated from the difference between electron and hole masses $\alpha$ and also, topological characteristics $\beta$. The strong spin-orbit coupling leads to distinct spin splitting at the valence band for different valleys. In the presence of an exchange field $h$ and superconducting gap induced by proximity effect, the Dirac-Bogoliubov-de Gennes (DBdG) Hamiltonian is given by:
\begin{equation}
\hhh=\left(\begin{array}{cc}
h_{0}-E_{F}+U(x)-sh&\Delta_{S}\\
\Delta_{S}^{\ast}&-h_{0}+E_{F}-U(x)-sh
\end{array}\right),
\end{equation}
where $h_{0}=\hbar v_{F}\mathbf{k}\cdot\bsigma_{\tau}+\Delta\sigma_{z}+\lambda s\tau(1-\sigma_{z})+\frac{\hbar^{2}\left|k\right|^{2}}{2m_{0}}(\frac{\alpha}{2}+\frac{\beta}{2}\sigma_{z})$, and $\bsigma_{\tau}=(\tau\sigma_{x},\sigma_{y})$ are the Pauli matrices. The spin-up and spin-down is labeled by $s=\pm 1$, and valley index $\tau=\pm 1$ denotes the $K$ and $K'$ valleys. The bare electron mass is $m_0=0.05\times 10^{-10}\:(eV s^2/m^2)$, and topological and mass difference band parameters are evaluated by $\beta=2.21$ and $\alpha=0.43$, respectively. $\Delta$ is the direct band gap, $\lambda\approx 0.04\:eV$ and $v_{F}=0.53\times 10^{6}\:m/s$ denote the spin-orbit coupling and Fermi velocity, respectively. The electrostatic potential $U(x)$ gives the relative shift of Fermi energy in $F$ and $S$ regions. The superconducting gap is presented by $\Delta_S$, which in the $d$-wave order parameter case, as mentioned in the previous section it is parameterized by the electron incidence angle (with respect to the perpendicular direction to the interface) $\theta_s$ in S region and orbital rotated angle $\alpha_s$, respectively. Taking the superconducting gap to be zero in F region and from the Hamiltonian Eq. (1), the excitation energy (relative to the Fermi energy $E_{FN}$) can be obtained as below:
\begin{equation}
\epsilon=\left|\lambda s \tau+\frac{\hbar^{2}\left|k\right|^2}{2m_0}(\frac{\alpha}{2})-E_{FN}-sh\pm\sqrt{\left(\Delta-\lambda s \tau+\frac{\hbar^{2}\left|k\right|^2}{2m_0}(\frac{\beta}{2})\right)^2+v^{2}_{F}\hbar^{2}\left|k\right|^{2}}\right|.
\end{equation}
The electrostatic potential $U(x)$ is determined to be zero in this region. Fermi energy is determined by the magnitude of the chemical potential. As illustrated in Fig. 1(a), the valence and conduction bands are characterized by two branches $\pm$ of the spectrum. Filled circle indicates the electron excitations, while empty circle denotes the hole excitations. The corresponding Fermi wavevector in ferromagnetic $MoS_2$ can be acquired from this eigenstates:
\begin{equation}
k_{F}=\left(\frac{\omega_{1}\frac{\alpha}{2}+\omega_{2}\frac{\beta}{2}+m_{0}v^{2}_{F}-\sqrt{(\omega_{1}\frac{\beta}{2}+\omega_{2}\frac{\alpha}{2})^2+(m_{0}v^{2}_{F})^2+2m_{0}v^{2}_{F}(\omega_{1}\frac{\alpha}{2}+\omega_{2}\frac{\beta}{2})}}{\frac{\hbar^2}{2m_{0}}(\frac{\alpha}{2}^2-\frac{\beta}{2}^2)}\right)^{1/2},
\end{equation}
where we define $\omega_{1}=-E_{FN}-\lambda s \tau+sh,\ \ \ \omega_{2}=-\lambda s \tau+\Delta$. The electron and hole excitations are indicated with states above and below the Fermi level, respectively. For $x>0$, the exchange field and electrostatic potential are taken to be zero and $-U_{0}$, respectively. Note that the mean-field conditions are satisfied as long as $\Delta_{0}\ll U_{0}+E_{FN}$. The dispersion relation of DBdG for electron-hole excitations is given by solving the energy-momentum quartic equation, as follows:
\begin{equation}
\epsilon_{sup.}=\pm\sqrt{\left|\Delta_S\right|^{2}+\left(\lambda s \tau+\frac{\hbar^{2}\left|k^{s}\right|^{2}}{2m_0}(\frac{\alpha}{2})-E_{FS}\pm\sqrt{\left(\Delta-\lambda s \tau+\frac{\hbar^{2}\left|k^{s}\right|^{2}}{2m_0}(\frac{\beta}{2})\right)^2+v_{F}^{2}\hbar^{2}\left|k^{s}\right|^{2}}\right)^{2}}.
\end{equation}
The wavevector and Fermi energy of superconducting quasiparticles are defined by $k^s$ and $E_{FS}(E_{FS}=E_{FN}+U_{0})$, respectively. The schematic of the above ML-MDS superconducting dispersion is shown in Fig. 1(b). We observe that the dispersion energy is strongly sensitive to Fermi energy for small values of wavevectors. For excitations below the superconducting gap there are no propagating waves in the superconductor. 

Hamiltonian (1) can be solved to obtain the wave function for two regions. Denoting the amplitude of normal and Andreev reflections, respectively, by $r$ and $r_{A}$, the incidence and reflected quasiparticle wavefunctions in the F section can be described by: 
$$
\psi_{F}=\frac{1}{\sqrt{\CN_{e}}}\left(\begin{array}{cc}
1\\
\tau e^{i\tau\theta_{e}} A_{Fe}\\
0\\
0
\end{array}\right)e^{i(\tau k_{x}x+k_{y}y)}+\frac{r}{\sqrt{\CN_{e}}}\left(\begin{array}{cc}
1\\
-\tau e^{-i\tau\theta_{e}} A_{Fe}\\
0\\
0
\end{array}\right)e^{i(-\tau k_{x}x+k_{y}y)}+
$$
\begin{equation}
+\frac{r_{A}}{\sqrt{\CN_{h}}}\left(\begin{array}{cc}
0\\
0\\
1\\
\tau e^{-i\tau\theta_{h}} A_{Fh}
\end{array}\right)e^{i(\tau k_{x}x+k_{y}y)},
\end{equation}
where we define $A_{Fe(h)}=\hbar v_{F} \left|k\right|/\left((-)\epsilon-E_{FN}+\Delta-2\lambda s \tau-\frac{\hbar^{2}\left|k\right|^{2}}{2m_0}(\frac{\alpha}{2}-\frac{\beta}{2})+sh\right)$. The normalization factor $\CN_{e(h)}$ ensure that the particle current density of states is the same. The charge and current density of quasiparticles may be defined by nonrelativistic and relativistic terms based on the Lorentz covariant continuity equation. Using the modified Dirac Hamiltonian, continuity equation results in:
$$
\frac{\partial}{\partial t}\left(\psi^{\dagger}\psi\right)+\frac{\partial}{\partial x}\left[v_{F}\left(\psi^{\dagger}(\tau\sigma_{x})\psi\right)+\frac{\hbar}{4im_{0}}\left(\psi^{\dagger}(\alpha+\beta\sigma_{z})\frac{\partial}{\partial x}\psi-\frac{\partial}{\partial x}\psi^{\dagger}(\alpha+\beta\sigma_{z})\psi\right)\right]=0.
$$
Hence, the normalization factor is given by:
\begin{equation}
\CN_{e(h)}=A_{Fe(h)}\cos(\tau\theta_{(h)})+\frac{\tau\hbar \left|k\right|}{4m_{0}v_{F}}\left((\alpha+\beta)+A^{2}_{Fe(h)}(\alpha-\beta)\right)\cos(\tau\theta_{(h)}).
\end{equation}

Inside the superconducting region, the solutions of the DBdG equation take the following more simple and accurate form:
\begin{equation}
\psi_{S}=t\left(\begin{array}{cc}
\zeta_{1}e^{-i\beta_{1}}\\
\zeta_{2}e^{-i\beta_{1}}e^{i\tau\theta_s}\\
e^{-i\gamma_{1}}\\
e^{-i\gamma_{1}}e^{i\tau\theta_s}
\end{array}\right)e^{i(\tau k_{x}^{s}x+k_{y}y)}+t'\left(\begin{array}{cc}
\zeta_{1}e^{i\beta_{2}}\\
-\zeta_{2}e^{i\beta_{2}}e^{-i\tau\theta_s}\\
e^{-i\gamma_{2}}\\
-e^{-i\gamma_{2}}e^{-i\tau\theta_s}
\end{array}\right)e^{i(-\tau k_{x}^{s}x+k_{y}y)},
\end{equation}
where the parameters $\zeta_{1(2)}$, $\gamma_{1(2)}$ and $\beta_{1(2)}$ can be expressed as:
$$
\zeta_{1(2)}=\left[\frac{(-)\left(-\Delta+\lambda s \tau-\frac{\hbar^{2}\left|k^{s}\right|^{2}}{2m_0}(\frac{\beta}{2})\right)}{\hbar v_{F}\tau k^{s}}-\sqrt{\left(\frac{\Delta-\lambda s \tau+\frac{\hbar^{2}\left|k^{s}\right|^{2}}{2m_0}(\frac{\beta}{2})}{\hbar v_{F}\tau k^{s}}\right)^2+1}\right]^{-1},
$$
$$
e^{i\gamma_{1(2)}}=\frac{\Delta_{S}^{e(h)}}{\left|\Delta_{S}^{e(h)}\right|} \ ; \ \ \ \beta_{1(2)}=\cos^{-1}{\left(\epsilon_{sup.}/\Delta_{S}^{e(h)}\right)}.
$$
The coefficients $t$ and $t'$ correspond to the transmission of electron and hole, respectively. By the subgap tunneling condition $\epsilon_{sup.}<\Delta_{0}$, the amplitude of the electron-hole conversion $r_{A}$ and electron-electron normal reflection $r$ can be found by the boundary condition at the interface between F and S regions:
\begin{equation}
r_{A}=\left(\frac{\CN_{e}}{\CN_{h}}\right)\frac{2\tau A_{Fe} \cos{(\tau\theta_{e})}}{\eta_{1}\eta_{4}e^{-i\beta_{1}}+\eta_{2}\eta_{3}e^{i\beta_{2}}}\left[\eta_{4}e^{-i\gamma_{1}}-\eta_{3}e^{-i\gamma_{2}}\right],
\end{equation}
\begin{equation}
r=\frac{2\zeta_{1}\tau A_{Fe} \cos{(\tau\theta_{e})}}{\eta_{1}\eta_{4}e^{-i\beta_{1}}+\eta_{2}\eta_{3}e^{i\beta_{2}}}\left[\eta_{4}e^{-i\beta_{1}}-\eta_{3}e^{i\beta_{2}}\right]-1,
\end{equation}
where
$$
\eta_{1(2)}=\zeta_{2}e^{(-)i\tau\theta_{s}}+(-)\zeta_{1}\tau A_{Fe}e^{-i\tau\theta_{e}},
$$
$$
\eta_{3(4)}=\tau A_{Fh}e^{-i\tau\theta_{h}}-(+)e^{i\tau\theta_{s}}.
$$
Finally, the tunneling conductance $G (eV)$ passing through the $F/S$ (or $N/S$, for the exchange field $h$ being zero) junction can now be calculated in terms of normal and Andreev reflection coefficients $r$ and $r_{A}$:
\begin{equation}
G(eV)=\sum_{s,\tau=\pm 1} G^{s,\tau}_{0}\int^{\theta_{c}}_{0}\left(1-\left|r\right|^{2}+\left|r_{A}\right|^{2}\right)\cos{\theta_{e}}d\theta_{e},
\end{equation}
where $G^{s,\tau}_{0}=e^{2}N_{s,\tau}(eV)/h$ is the ballistic conductance of spin and valley-dependent transverse modes $N_{s,\tau}=kw/\pi$ in a sheet of $MoS_{2}$ of width $w$ that $eV$ denotes the bias voltage. The upper limit of integration in Eq. (10) needs to obtain exactly based on the fact that the incidence angle of electron-hole in the two regions must be less than $\pi/2$, where we find the critical angle $\theta_{c}=\arcsin k^{h}/k^{e}$ to take place the actual Andreev reflection.

\section{Andreev reflection}

First we consider the scattering process in N/S junction (no ferromagnetic exchange field $sh=0$) on top of ML-MDS to obtain precisely the reflection of charge carriers versus incidence angle to the interface, since we may fix the range of chemical potential to remain in the valence band of either normal or ferromagnetic section. It confirms that the subgap transmission of electrons with excitations below the superconducting gap is forbidden (note that, $\left|r\right|^{2}+\left|r_A\right|^{2}=1$). We know that at the normal incidence, the Andreev reflection happens with unit probability in Dirac materials.
In Fig. 2, obviously, Andreev and normal reflections of Eqs. (8) and (9) versus incidence angle are demonstrated. The spin-valley polarized Andreev reflection for incident electron from the left normal region with spin $s$ from valley $\tau$ extremely depends on the Fermi level of valence band. Therefore, our calculations should be done under the condition that the Fermi energy of normal section satisfies the relation $-\Delta-2\lambda <E_{FN}<-\Delta+2\lambda$. The magnitude of Fermi energy in this structure is determined by the incident electron from $K$ valley with spin-down $(E_{FN}=-1.03\ eV)$. As expected, the pure symmetric reflection curves of Andreev and normal are found. Considering the doping of superconducting section reveals that for both $n$ and $p$-doped cases, the electron-hole conversion happens with most efficiency at normal incidence. It is seen from Fig. 2 that the AR amplitude decreases slowly and goes to zero at $\theta_{e}=\pi/2$ for $n$-doped case, while in $p$-doped case the AR curve behaves as decreasing for smaller incidence angles. We obtain the unit AR at the normal incidence for $n$-doped case, which is not observed in the previous work $\cite{MRA}$.
As long as the condition $-1.03<E_{FN}<-0.87$ is satisfied, the magnitude of Fermi energy in normal section affects the probability of AR. Accordingly, we plot the $\theta_{e}$ dependence of AR in Fig. 3 for different values of Fermi energy of normal section. From Fig. 3, by increasing the absolute Fermi energy via chemical potential we can increase the AR probability. It is interesting that the electron-hole conversion with unit probability happens at maximum value of allowed absolute Fermi energy of normal segment. 

In the next step, we investigate how the asymmetric $d$-wave superconducting gap can affect fundamentally the reflections behavior, which is shown in Fig. 4. In this case, the magnitude of AR decreases comparing with those in the $s$-wave. This is a consequence of angular averaging in the unconventional pairing symmetry. Importantly, relative to the superconducting orbital rotated angle $\alpha\in[0-\pi/4]$, the AR suddenly falls for a specific incidence angle. So, the behavior of the Andreev reflection is not a uniform curve in the presence of $d$-wave pairing potential.
An interesting aspect is when an F metal is included in normal section. It is notable that in presence of exchange field $sh$, spin-splitting of valence band of $MoS_2$ strongly depends on the valley index. As the result, an extra momentum change between Andreev reflected electron and hole occurs, and actually it leads to decrease the amplitude of AR. By increasing the induced exchange field, the separation between spin subbands decreases in $K$ valley, and reaches to zero for $h=2\lambda$, whereas increases in $K'$ valley. Consequently, it needs to define the Fermi energy (chemical potential) in the F/S structure. Since, there are two inequivalent valleys in ML-MDS, we have four critical Fermi energy for different spin subbabnds of two valleys as ($-\Delta+2\lambda-h$), ($-\Delta+2\lambda+h$), ($-\Delta-2\lambda-h$) and ($-\Delta-2\lambda+h$). Actually, the spin-valley polarized Andreev reflection occurs when we have such condition as $-\Delta+2\lambda-h<E_{FN}<-\Delta+2\lambda+h$. Thus, similar to N/S structure, AR probability increases by increasing the absolute Fermi energy. Furthermore, only the incoming spin up quasiparticle from $K$ valley can be reflected as a hole state with opposite spin and different valley index. Otherwise, there is no longer a hole state in the F region, and Andreev process is suppressed. In the numerical results, the exchange field is taken in units of the spin-orbit coupling $h=m\lambda (m<2)$. The resulting zero bias Andreev and normal reflections curves are presented in Figs. 5(a) and (b) for two different magnitudes of exchange filed. The results of $s$-wave order parameter are displayed in Fig. 5(a). It is shown that under the effect of exchange energy, the AR amplitude decreases. Interestingly, for $d$-wave superconducting gap it declines impressively.

\section{Conductance}

We now turn to a study of the tunneling conductance in the ML-MDS $N/S$ and $F/S$ structures. From Eq. (2) as long as we have $\epsilon\leq E_{F}$, the Andreev reflected hole is in the valence band, and AR may be retro. From Fig. 6, it can be seen that the magnitude of subgap Andreev conductance versus bias voltage is sensitive to Femi energy. Depending on the magnitude of gate voltage, the enhanced conductance can occur for subgap bias voltages. It is shown that the zero bias conductance increases with Fermi energy of superconductor section $E_{FS}$. Also, increasing the AR probability with $E_{FN}$ leads to the increasing the resulting Andreev conductance. This behavior can be exactly explained in the framework of the Andreev process occurring in the previous section.
In order to explain the behavior of the Andreev conductance for different superconducting gap, we demonstrate the Fermi energy dependence of the Andreev conductance, see Fig. 7. As a result, the conductance-Fermi energy relation for the $s$-wave and anisotropic $d$-wave is qualitatively the same, although its magnitude is reduced in $d$-wave gap, which is related to the effective weakening of the gap upon Fermi surface averaging of its absolute value compared to the $s$-wave. As shown in Fig. 7, for larger mismatch between electron and hole wavevectors, the difference between conductances of various order parameters is obvious.

Finally, we investigate how the magnetization of F region can affect the Andreev conductance. The incoming quasiparticle from the F region can include various combinations of valley and spin indices. It is noticeable that only the incident electron with spin-up from valley $K$ can be reflected as hole. For other quasiparticles, the AR process is suppressed. Accordingly, the Andreev conductance in ferromagnetic structure is significantly decreased. The behavior of zero bias conductance versus bias voltage for three different values of exchange field is presented in Fig. 8. As seen from Figs. 5, in presence of exchange field the probability of AR diminishes in zero bias. So, the resulting Andreev conductance decreases with $h$. We find that the exchange energy can decline the subgap Andreev conductance in $MoS_{2}$-based $F/S$ junction in contrast to the graphene-based junction.

\section{Conclusion}

In summary, the Andreev process and transport characteristics of monolayer $MoS_2$ $F/S$ junction have been studied. Using the modified Dirac Hamiltonian, that in addition to the first order term of momentum contains the quadratic terms corresponding to the topological and difference mass between electron and hole terms, we have given the explicit form of electron-hole superconducting excitations and resulting Dirac spinors. Relative to the effective spin-orbit coupling strongly appeared in valence band of ferromagnetic ML-MDS, the allowed values of chemical potential in order to have subgap conductance have been fixed. Note that, the Andreev reflection of an incidence electron from the left ferromagnetic region with spin number $s$ from valley $\tau$ extremely depends on its Fermi level in the valence band. We have found the AR to exhibit a new behavior with respect to the ferromagnetic exchange field and superconducting order parameter. In particular, the unit value of AR for normal incidence has been obtained in both $n$- and $p$-doped superconductor $MoS_2$. Using the obtained compact form of superconducting states, we have investigated the role of singlet anisotropic superconducting gap in the Andreev process and resulting subgap conductance. Considering the dynamical feature of monolayer $MoS_2$, the Andreev-Klein transmission between inequivalent valleys can happen only in the valence band, and resulting retro AR gives rise to decline the valley-resolved Andreev conductance in the presence of exchange field. Finally, the anisotropic $d$-wave superconducting gap causes the breaking of AR curve, and resulting zero bias conductance enhances with increasing the absolute of Fermi energy controlled by the spin-valley polarized AR.  

%\begin{flushleft}
%\textbf{Acknowledgment}
%\end{flushleft}
%The authors would greatly thank the anonymous reviewers for useful comments, which have greatly helped us in improving this paper.

\newpage
\textbf{Figure captions}\\
\textbf{Figure 1(a), (b)} (Color online) (a) the energy dispersion in momentum space at the ML-MDS. Red and blue curves indicate spin up and down subbands in valence band, respectively. Black and dashed line distinguishes the conductance band. Filled states above the Fermi energy indicate electron excitations, while empty states below the Fermi level indicate hole excitations, (b) The energy dispersion in superconductor ML-MDS, calculated from Eq. (4).\\
\textbf{Figure 2} (Color online) Plot of the probability of normal reflection $(\left|r\right|^{2})$ and AR $(\left|r_{A}\right|^{2})$ as a function of the incidence angle in the $s$-wave case with n-doped S region (solid lines) and p-doped S region (dashed lines) when $E_{FN}=-1.03\ eV$ and $\epsilon=0$.\\
\textbf{Figure 3} (Color online) Probability of the AR as a function of incident angle for several values of normal Fermi energy when $E_{FS}=15\ eV$ and $\epsilon=0$.\\
\textbf{Figure 4} (Color online) Plot of the probability of normal reflection and AR in the $d$-wave case for two values of rotated angle $\alpha$. We have set $E_{FN}=E_{FS}=-1.03\ eV$ and $\epsilon=0$.\\
\textbf{Figure 5(a), (b)}(Color online) Behavior of the probabilities of normal and Andreev reflections versus incidence angle in presence of exchange field for (a) $s$-wave and (b) $d$-wave order parameter, when $E_{FS}=2\ eV$, $\epsilon=0$ and $\alpha=0.2\ \pi$.\\
\textbf{Figure 6} (Color online) Normalized Andreev conductance as a function of the bias voltage for different magnitude of Fermi energy. We have set $\Delta_{S}=0.02\ eV$.\\
\textbf{Figure 7} (Color online) Behavior of the conductance for $s$ and $d$-wave superconducting gap versus normal Fermi energy. The magnitude of orbital rotated angle is $\alpha=0.2\ \pi$.\\
\textbf{Figure 8} (Color online) Andreev conductance of the ML-MDS-based F/S structure as a function of bias voltage for different values of the exchange field in F region. We obtain maximum value for conductance when $E_{FN}=-0.99,-0.97,-0.96$ for $m=1,1.5,1.7$ respectively. We have set $E_{FS}=2\ eV$ and $\Delta_{S}=0.02\ eV$.\\

\newpage

\begin{figure}[p]
\epsfxsize=0.5 \textwidth
\begin{center}
\epsfbox{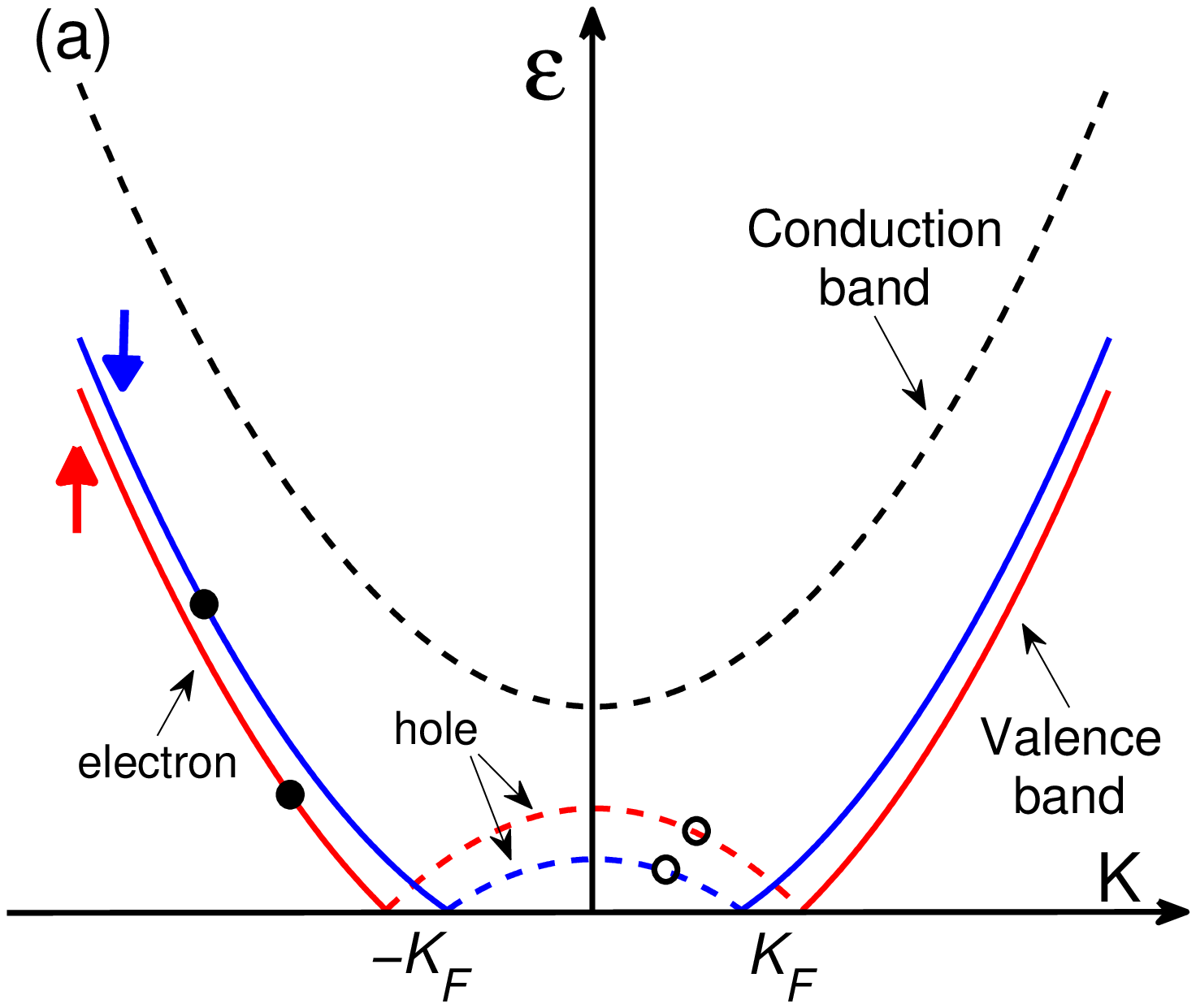}
%\setcounter{figure}{0}
%\caption{\footnotesize (a)}
\end{center}
\end{figure}

\begin{figure}[b]
\epsfxsize=0.6 \textwidth
\begin{center}
\epsfbox{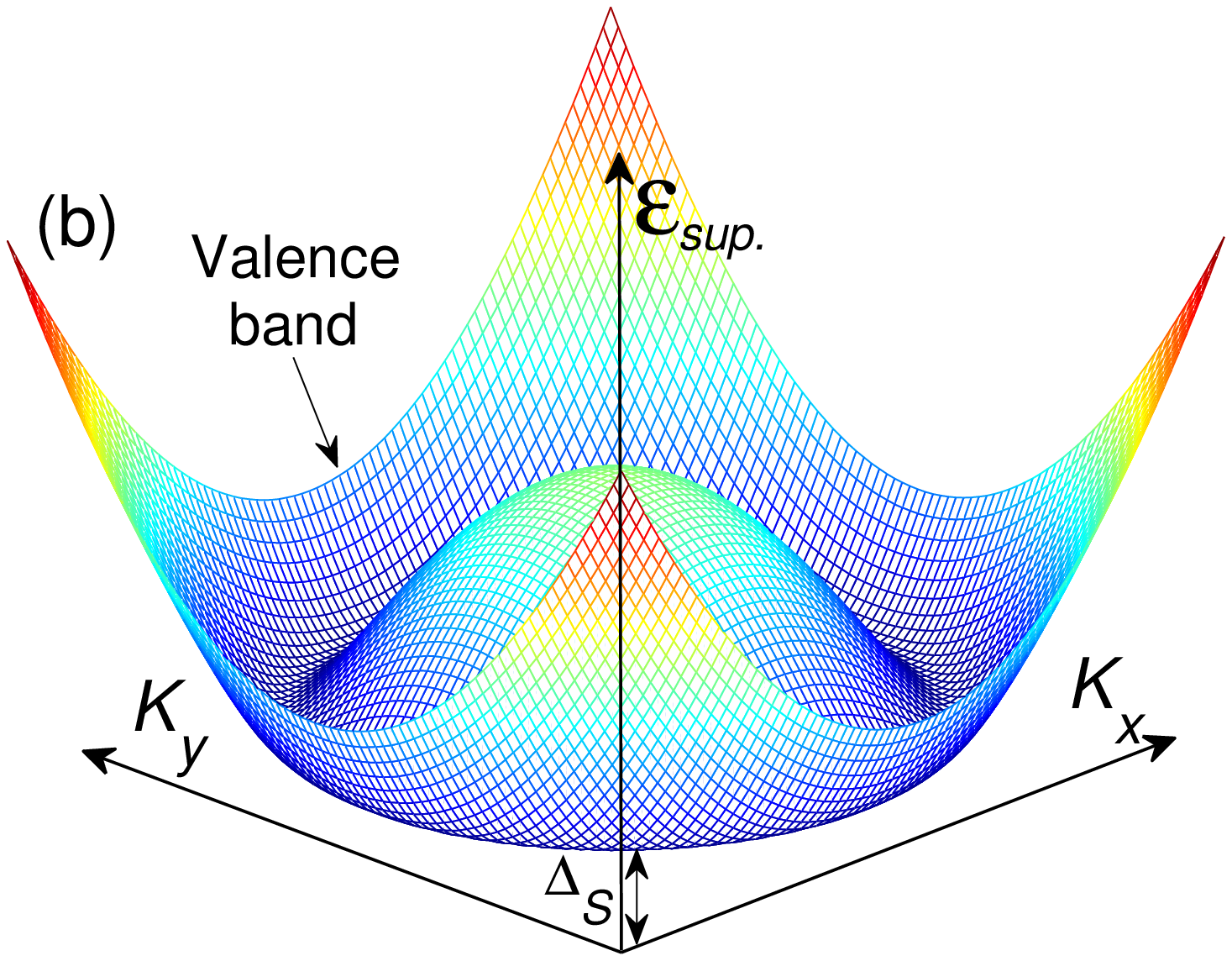}
\setcounter{figure}{0}
\caption{\footnotesize (a), (b)}
\end{center}
\end{figure}

\begin{figure}[b]
\epsfxsize=0.6 \textwidth
\begin{center}
\epsfbox{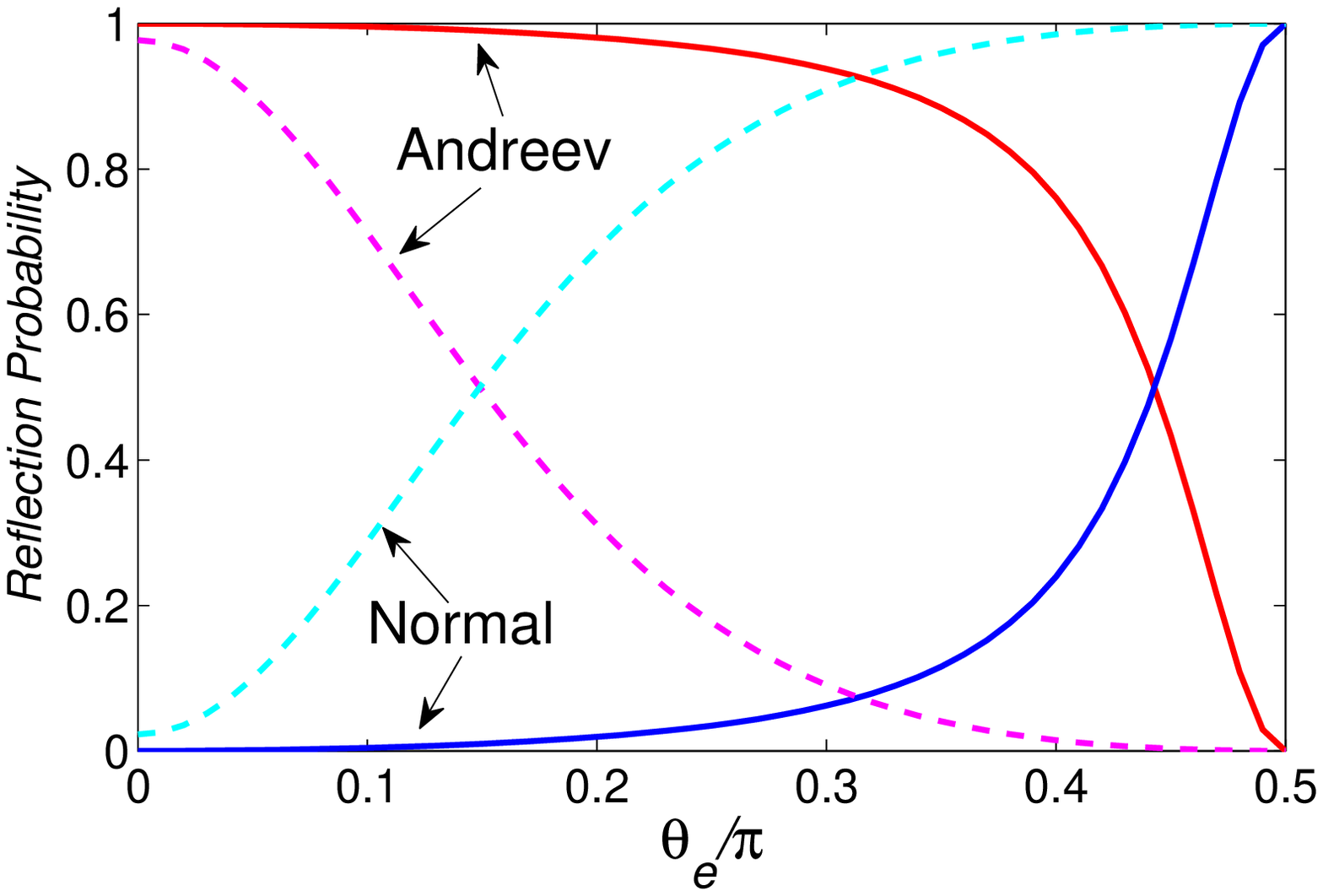}
\setcounter{figure}{1}
\caption{\footnotesize }
\end{center}
\end{figure}

\begin{figure}[p]
\epsfxsize=0.6 \textwidth
\begin{center}
\epsfbox{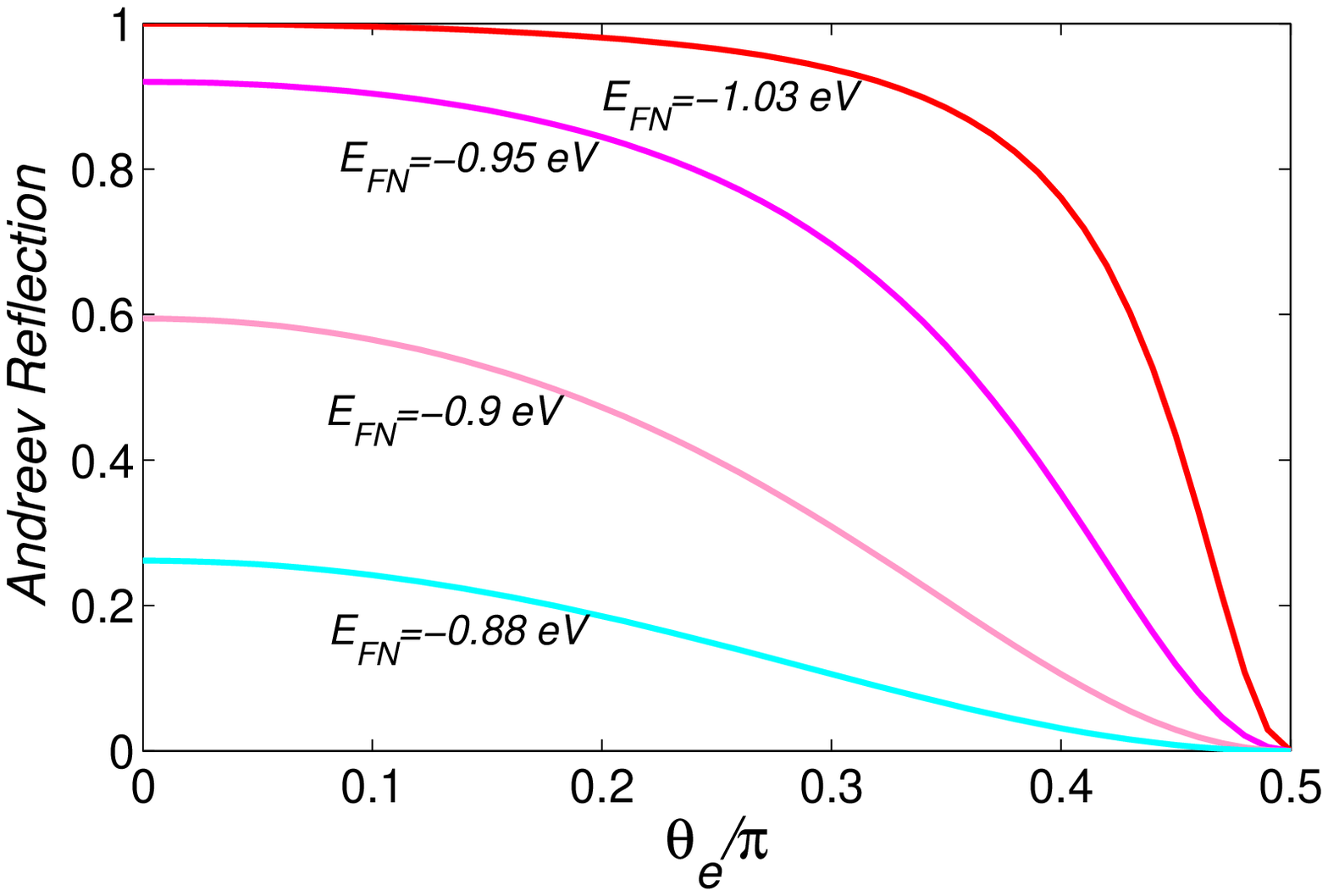}
\setcounter{figure}{2}
\caption{\footnotesize }
\end{center}
\end{figure}

\begin{figure}[p]
\epsfxsize=0.6 \textwidth
\begin{center}
\epsfbox{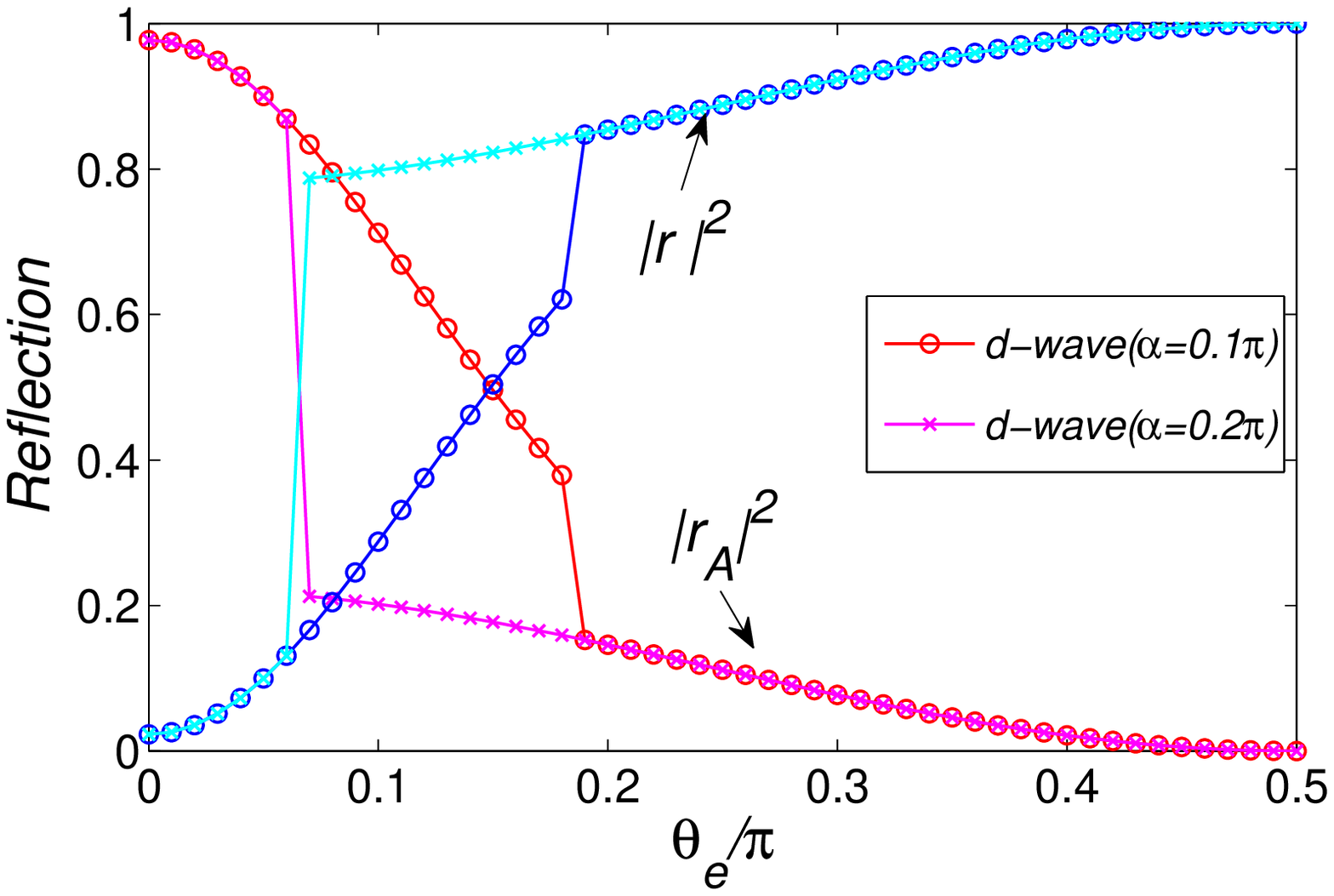}
\setcounter{figure}{3}
\caption{\footnotesize }
\end{center}
\end{figure}

\begin{figure}[p]
\epsfxsize=0.8 \textwidth
\begin{center}
\epsfbox{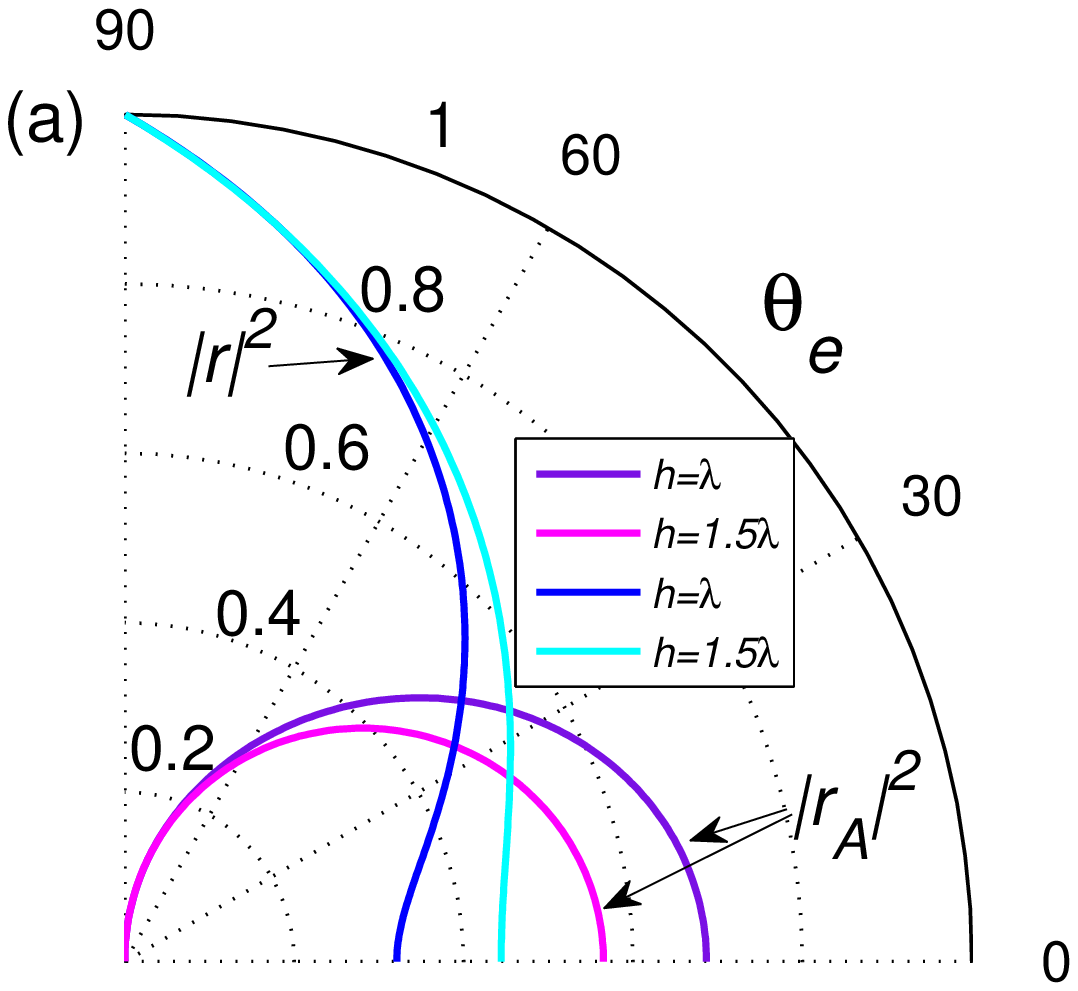}
%\setcounter{figure}{4}
%\caption{\footnotesize }
\end{center}
\end{figure}

\begin{figure}[p]
\epsfxsize=0.8 \textwidth
\begin{center}
\epsfbox{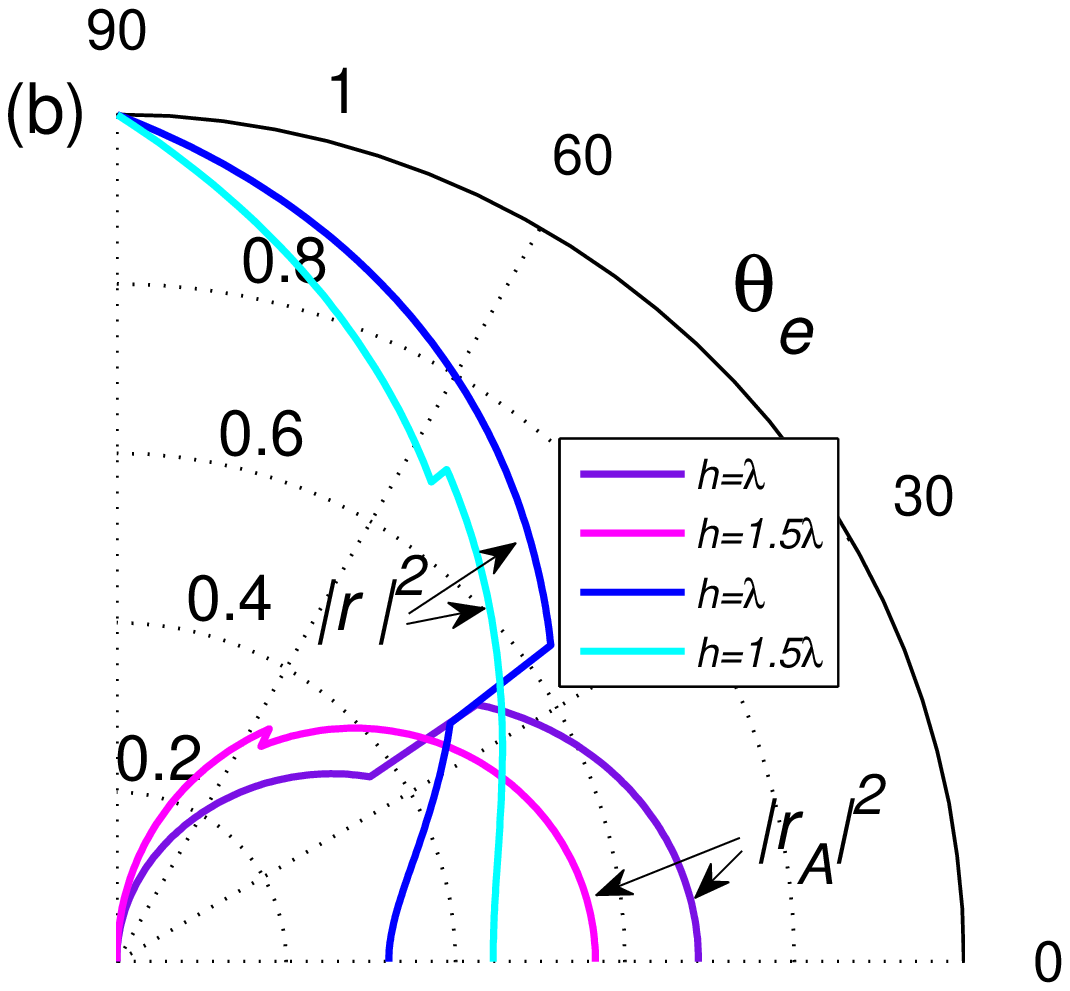}
\setcounter{figure}{4}
\caption{\footnotesize (a), (b)}
\end{center}
\end{figure}

\begin{figure}[p]
\epsfxsize=0.6 \textwidth
\begin{center}
\epsfbox{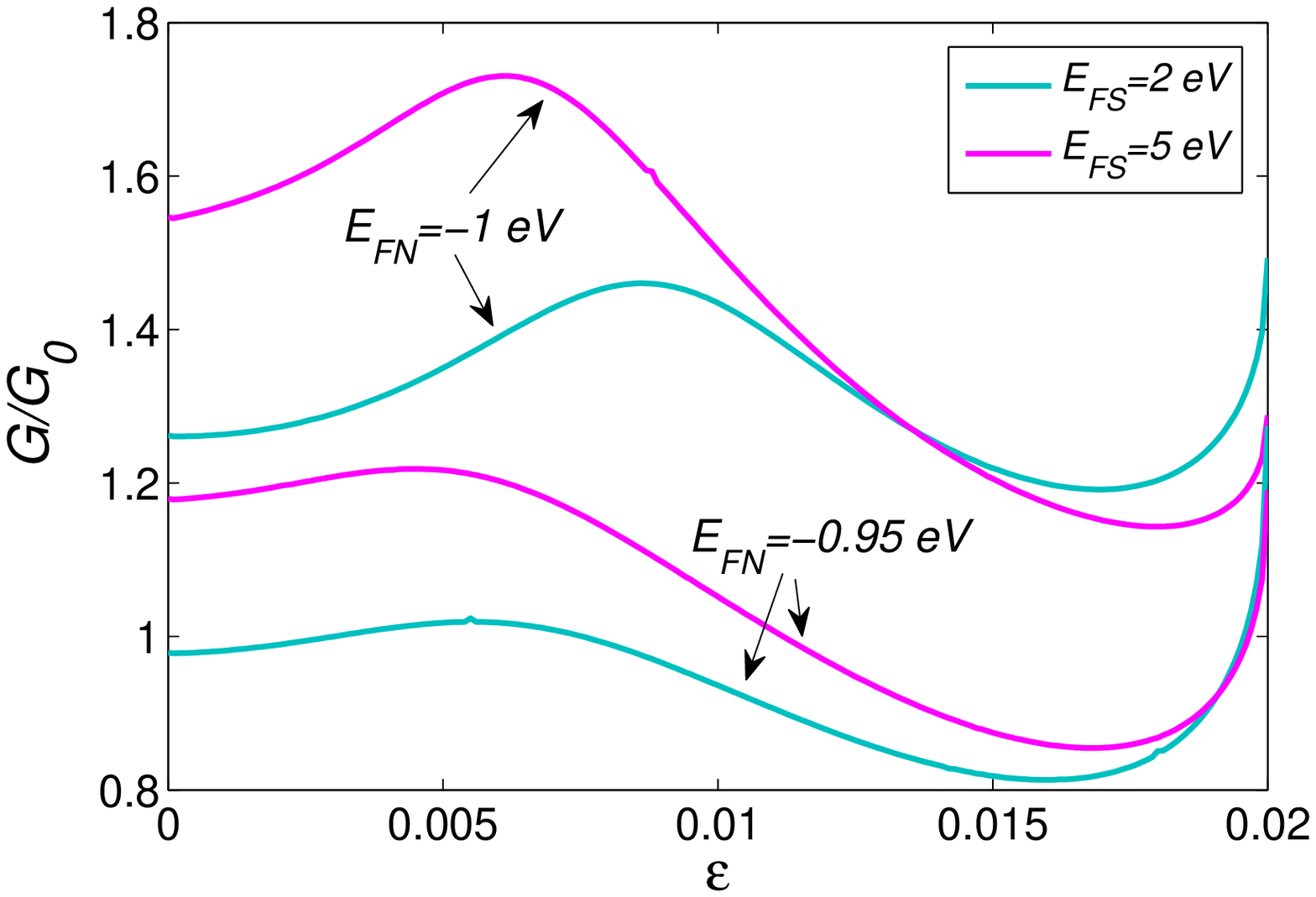}
\setcounter{figure}{5}
\caption{\footnotesize }
\end{center}
\end{figure}

\begin{figure}[p]
\epsfxsize=0.6 \textwidth
\begin{center}
\epsfbox{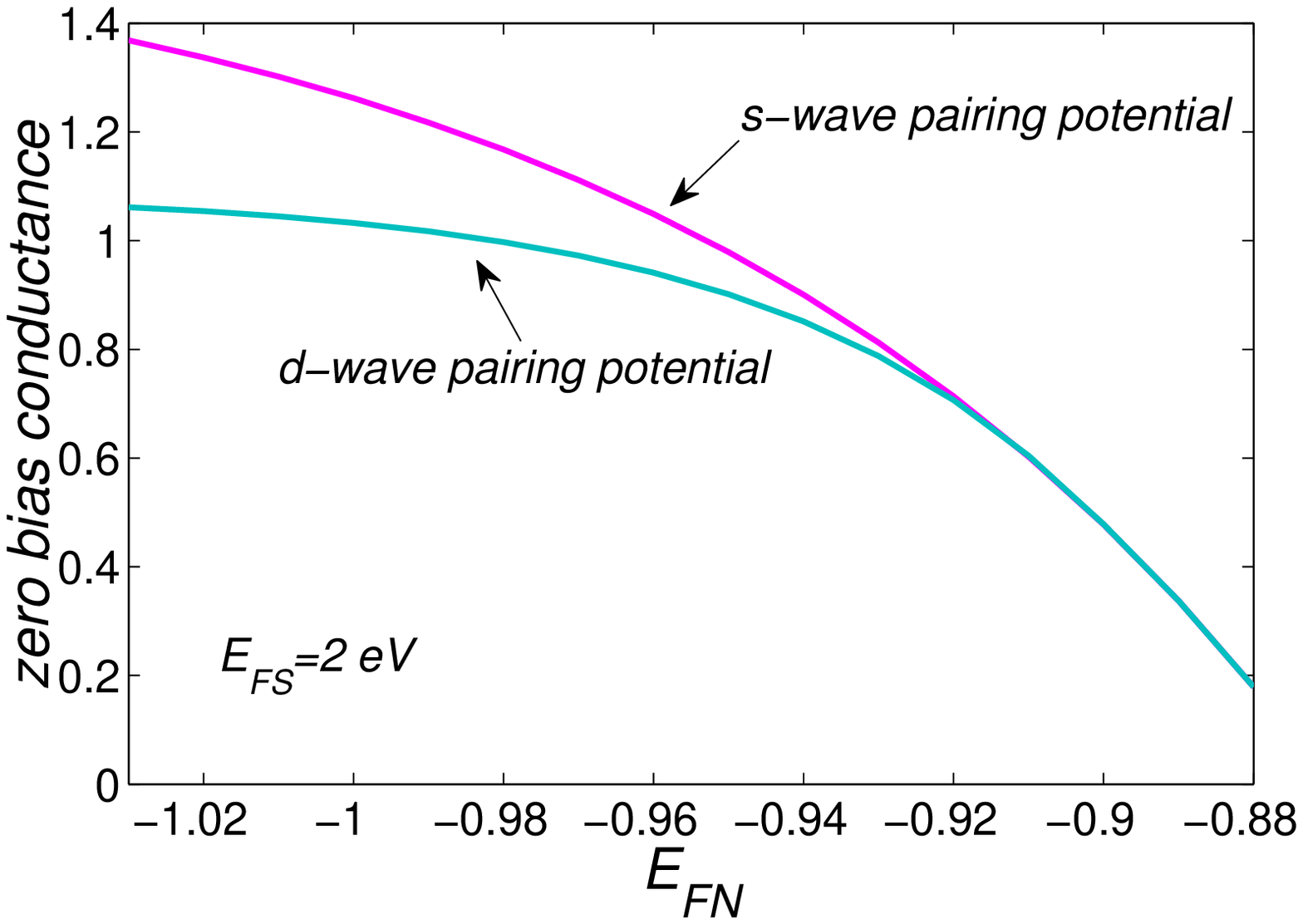}
\setcounter{figure}{6}
\caption{\footnotesize }
\end{center}
\end{figure}

\begin{figure}[p]
\epsfxsize=0.6 \textwidth
\begin{center}
\epsfbox{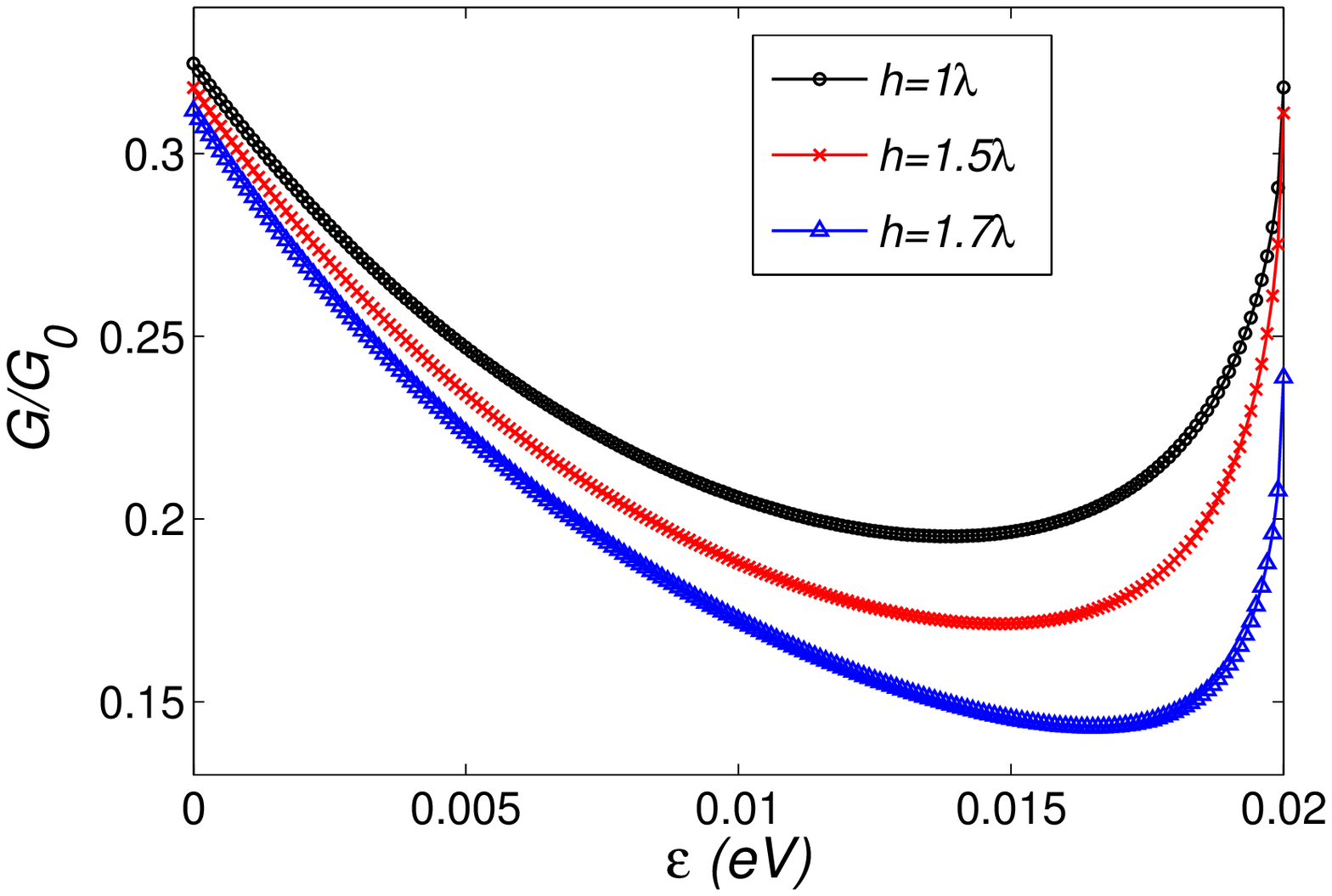}
\setcounter{figure}{7}
\caption{\footnotesize }
\end{center}
\end{figure}

\end{document}